# A note on an absorption problem for a Brownian particle moving in a harmonic potential

Michael J. Kearney [1] and Richard J. Martin [2]

[1] Senate House, University of Surrey, Guildford, Surrey, GU2 7XH, UK

[2] Department of Mathematics, Imperial College London,

South Kensington, London, SW7 2AZ, UK

*Abstract*

An analysis is presented of a Brownian particle moving on the half-line, subject to a restoring force proportional to its displacement and an absorbing boundary at the origin. When the initial displacement is large, the central moments of the time to be absorbed tend to finite constants, as do the position moments when evaluated at the most probable absorption time. These quantities are derived explicitly.

1. Introduction

Consider an overdamped Brownian particle whose position or displacement, in suitable units, evolves according to the Langevin equation

$$\frac{dx}{dt} = -x + \sqrt{2}\xi(t). \qquad (1)$$

This describes motion in a harmonic potential $V(x) = \frac{1}{2}x^2$, equating to an external force $F(x) = -x$ which drives the particle towards zero, whilst $\xi(t)$ is a delta-correlated Gaussian noise term. Without noise, the particle relaxes according to $x(t) = \alpha e^{-t}$, where $\alpha$ is the initial position. With noise, the particle can reach zero in finite time, at which point the assumption is made that it is instantly absorbed. Given $\alpha > 0$, the motion is therefore constrained to the half-line $[0, \infty)$.

A question of interest concerns the statistics of this random time to absorption or first hitting time, namely $\mathcal{T} \equiv \min\{t : x(t) = 0 | \alpha\}$ with $x(t \geq \mathcal{T}) = 0$. The purpose of this note is to analyse what happens when the particle starts far away from zero. Somewhat counterintuitively, the central moments of $\mathcal{T}$ reduce to finite constants in the limit $\alpha \to \infty$ [1]. This stems from an underlying invariance within the problem, made clear here by the method of derivation which is complementary to and more comprehensive than that provided in [1]. Additional insights beyond the scope of [1] come from studying the limiting behaviour of the moments of the particle's position when evaluated at the most probable absorption time.

## 2. Moments relating to the particle's time to absorption

Let $p(x,t|\alpha)$ for $x \geq 0$ be the particle position density at time $t$ given initial condition $p(x,t=0|\alpha) = \delta(x-\alpha)$ and with an absorbing boundary at $x=0$ such that $p(x=0,t|\alpha) = 0$. Based on (1), this satisfies a backward Fokker-Planck equation:

$$\left[\frac{\partial^2}{\partial \alpha^2} - \alpha \frac{\partial}{\partial \alpha}\right] p(x,t|\alpha) = \frac{\partial}{\partial t} p(x,t|\alpha) \tag{2}$$

with supplementary condition $p(x \to \infty, t|\alpha) = 0$. It follows that the zeroth moment of the particle position, namely $\mathrm{E}_t[1|\alpha] = \int_0^\infty p(x,t|\alpha)\mathrm{d}x$, satisfies

$$\left[\frac{\partial^2}{\partial \alpha^2} - \alpha \frac{\partial}{\partial \alpha}\right] \mathrm{E}_t[1|\alpha] = \frac{\partial}{\partial t} \mathrm{E}_t[1|\alpha]; \qquad \mathrm{E}_{t=0}[1|\alpha] = 1$$

with the following solution:

$$\mathrm{E}_t[1|\alpha] = \mathrm{erf}\left(\frac{\alpha e^{-t}}{\sqrt{2(1-e^{-2t})}}\right). \tag{3}$$

This decays with time since $p(x,t|\alpha)$ concerns sample paths which have not been absorbed. Accordingly, $\mathrm{E}_t[1|\alpha]$ is equivalent to the probability that the particle has not been absorbed before time $t$, i.e. $\mathrm{E}_t[1|\alpha] \equiv \Pr(\mathcal{T} > t)$. One therefore has

$$\mathrm{E}_t[1|\alpha] = \int_t^\infty P(\mathcal{T}|\alpha)\mathrm{d}\mathcal{T} \quad \Leftrightarrow \quad P(\mathcal{T}|\alpha) = -\frac{\partial}{\partial t}\mathrm{E}_t[1|\alpha]\bigg|_{t=\mathcal{T}}$$

from which one can obtain the density of $\mathcal{T}$:

$$P(\mathcal{T}|\alpha) = \sqrt{\frac{2}{\pi}}\frac{\alpha e^{-\mathcal{T}}}{(1-e^{-2\mathcal{T}})^{3/2}}\exp\left\{-\frac{\alpha^2 e^{-2\mathcal{T}}}{2(1-e^{-2\mathcal{T}})}\right\}. \qquad (4)$$

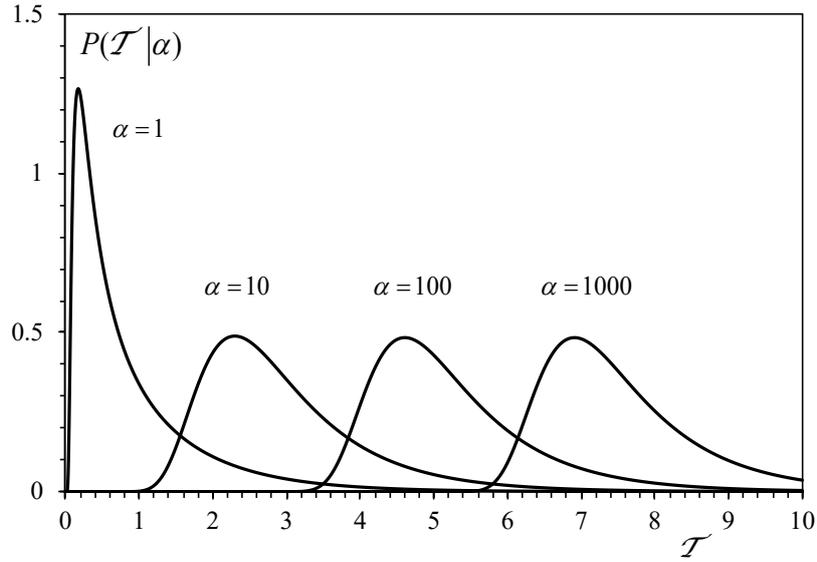

Figure 1. The absorption time density for different initial conditions.

This is well known from the theory of the Ornstein-Uhlenbeck process [2-4]. Example plots for representative values of $\alpha$ are shown in figure 1. The point at which the density is maximum (the most probable absorption time) is given by

$$\mathcal{T}_{\max}(\alpha) = \frac{1}{2}\log\left(\frac{\alpha^2 - 1 + \sqrt{\alpha^4 - 2\alpha^2 + 9}}{2}\right).$$

As $\alpha \to \infty$, this reduces to $\mathcal{T}_{max}(\alpha) = \log \alpha + O(\alpha^{-2})$. The logarithmic behaviour is evident in figure 1, and a clear invariance (up to translation) emerges in the density profile as $\alpha \to \infty$.

It is worth remarking that if the absorbing boundary is anywhere other than at zero, there is no exact solution in the sense of a simple generalisation of (4). An extensive literature is devoted to approximate treatments; see e.g. [5-8], with recent developments given in [9-11]. The general case, however, does not have the same interesting characteristics as $\alpha \to \infty$. Nor is it easy to deduce any of the findings below as a special case of the general results presented in [2, 3, 5-7].

Guided by figure 1, for large initial displacement one may propose the equality in distribution:

$$\mathcal{T} \stackrel{\mathcal{D}}{=} \log \alpha + \xi; \qquad \alpha \to \infty \tag{5}$$

where $\xi$ is a random variable, *independent* of $\alpha$, whose density follows from substituting (5) into (4) and simplifying in the given limit:

$$P(\xi) \equiv \lim_{\alpha \to \infty} P(\mathcal{T} = \log \alpha + \xi | \alpha) = \sqrt{\frac{2}{\pi}} \exp\left[-\xi - \tfrac{1}{2} e^{-2\xi}\right]. \tag{6}$$

This is a relative of the Gumbel distribution which occurs in extreme value theory. By way of characterisation one has

$$\Pr(\xi > 0) = \mathrm{erf}\left(\frac{1}{\sqrt{2}}\right) = 0.682...$$

$$\mathrm{E}[\xi] = \frac{1}{2}(\log 2 + \gamma) = 0.635...$$

where $\gamma = 0.577...$ is Euler's constant. It follows from (5) that as $\alpha \to \infty$:

$$\mathrm{E}[\mathcal{T}] = \log \alpha + \frac{\log 2 + \gamma}{2} + o(1). \tag{7}$$

This was derived a different way in [1].

To extend this to higher order moments, it is natural to consider the central moments, which are defined by $C_n(\alpha) \equiv \mathrm{E}\left[(\mathcal{T} - \mathrm{E}[\mathcal{T}])^n\right]$, with $C_0(\alpha) = 1$ and $C_1(\alpha) = 0$. The formal generating function $C(s, \alpha)$ may be written as

$$C(s, \alpha) \equiv \mathrm{E}[e^{s(\mathcal{T} - \mathrm{E}[\mathcal{T}])}] = \sum_{n=0}^{\infty} C_n(\alpha) \frac{s^n}{n!}.$$

It follows from (5, 6, 7) that

$$C(s, \infty) = \frac{2}{\sqrt{\pi}} \int_{-\infty}^{\infty} e^{s(\xi - \frac{1}{2}(\log 2 + \gamma))} \exp\left(-\xi - \tfrac{1}{2} e^{-2\xi}\right) d\xi = \frac{e^{-s\varphi/2}}{\sqrt{\pi}} \Gamma\left(\frac{1-s}{2}\right) \tag{8}$$

where $\varphi \equiv 2\log 2 + \gamma$. To extract $C_n(\infty) \equiv \mathrm{E}\left[(\xi - \tfrac{1}{2}(\log 2 + \gamma))^n\right]$ one can manipulate (8) to write

$$C_n(\infty) = \frac{\partial^n}{\partial s^n}\left[\frac{e^{-s\varphi/2}}{\sqrt{\pi}}\Gamma\left(\frac{1-s}{2}\right)\right]\bigg|_{s=0} = (-1)^n \frac{e^{-\varphi/2}}{\sqrt{\pi}\,2^n}\frac{\partial^n}{\partial z^n}\left[e^{z\varphi}\Gamma(z)\right]\bigg|_{z=\frac{1}{2}}$$

$$= (-1)^n \frac{e^{-\varphi/2}}{\sqrt{\pi}\,2^n}\frac{\partial^{n-1}}{\partial z^{n-1}}\left[(\varphi + \psi(z))e^{z\varphi}\Gamma(z)\right]\bigg|_{z=\frac{1}{2}}.$$

(9)

Here, $\psi(z)$ is the digamma function:

$$\psi(z) \equiv \frac{\mathrm{d}\log\Gamma(z)}{\mathrm{d}z} = -\gamma - \frac{1}{z} + \sum_{j=1}^{\infty}\left[\frac{1}{j} - \frac{1}{j+z}\right].$$

Defining $f(z) = \varphi + \psi(z)$ and $g(z) = e^{z\varphi}\Gamma(z)$, and noting that $f(\tfrac{1}{2}) = 0$, the expression (9) can be evaluated recursively:

$$C_n(\infty) = (-1)^n \frac{e^{-\varphi/2}}{\sqrt{\pi}\,2^n}\sum_{k=1}^{n-1}\binom{n-1}{k}\left\{\frac{\partial^k}{\partial z^k}f(z)\right\}\bigg|_{z=\frac{1}{2}}\left\{\frac{\partial^{n-1-k}}{\partial z^{n-1-k}}g(z)\right\}\bigg|_{z=\frac{1}{2}}$$

$$= \frac{1}{2^n}\sum_{k=1}^{n-1}\binom{n-1}{k}\times \psi^{(k)}(\tfrac{1}{2})\times (-1)^{-1-k}\,2^{-1-k}\,C_{n-1-k}(\infty)$$

where for $k \geq 1$ one has

$$\psi^{(k)}(z) \equiv \frac{d^k \psi(z)}{dz^k} = (-1)^{k+1} k! \sum_{j=1}^{\infty} \frac{1}{(j+z)^{k+1}}$$

$$\psi^{(k)}(\tfrac{1}{2}) = (-1)^{k+1} k! (2^{k+1} - 1) \zeta(k+1).$$

Here, $\zeta(z)$ is the Riemann zeta function. It follows for $n \geq 2$ that

$$C_n(\infty) = (n-1)! \sum_{k=2}^{n} \frac{(2^k - 1)}{2^k (n-k)!} \zeta(k) C_{n-k}(\infty). \tag{10}$$

The limiting central moments as $\alpha \to \infty$ are clearly finite constants, with the first non-trivial result being $C_2(\infty) = \pi^2 / 8$. Since $C_2(\alpha)$ is the variance, which is a strictly increasing function of $\alpha$, this is bounded above by $\pi^2 / 8$. Concerning the basic moments, one can write

$$E[\mathcal{T}^n] \equiv E\left[\left(E[\mathcal{T}] + (\mathcal{T} - E[\mathcal{T}])\right)^n\right] = \sum_{k=0}^{n} \binom{n}{k} E[\mathcal{T}]^k C_{n-k}(\alpha)$$

which together with (7) implies that as $\alpha \to \infty$:

$$E[\mathcal{T}^n] = \sum_{k=0}^{n} \binom{n}{k} \left(\log \alpha + \frac{\log 2 + \gamma}{2}\right)^k C_{n-k}(\infty) + o(1). \tag{11}$$

As shown in [1], one can derive a closed-form expression for the limiting cumulants $K_n(\infty)$. Here, an alternative approach is presented. Starting from the formal generating function $K(s, \alpha)$:

$$K(s,\alpha) \equiv \log \mathrm{E}[e^{s\mathcal{T}}] = s\mathrm{E}[\mathcal{T}] + \log C(s,\alpha) = \sum_{n=1}^{\infty} K_n(\alpha) \frac{s^n}{n!}$$

it follows based on (8) that

$$\lim_{\alpha \to \infty} \left[ K(s,\alpha) - s\mathrm{E}[\mathcal{T}] \right] = \log\left[ \frac{e^{-s\varphi/2}}{\sqrt{\pi}} \Gamma\left(\frac{1-s}{2}\right) \right]. \tag{12}$$

Using the gamma function duplication formula and the identity

$$\log \Gamma(1+z) = -\gamma z + \sum_{n=2}^{\infty} \frac{\zeta(n)}{n} (-z)^n$$

one can rewrite (12) as

$$\lim_{\alpha \to \infty} \left[ K(s,\alpha) - s\mathrm{E}[\mathcal{T}] \right] = -\left(\frac{\gamma}{2}\right) s + \log \Gamma(1-s) - \log \Gamma(1-\tfrac{1}{2}s)$$
$$= \sum_{n=2}^{\infty} \frac{\zeta(n)}{n}\left(1 - \frac{1}{2^n}\right) s^n. \tag{13}$$

The result for $K_n(\infty)$ follows by inspection:

$$K_n(\infty) = (n-1)!\left(1 - \frac{1}{2^n}\right)\zeta(n); \qquad n \geq 2 \tag{14}$$

and the first few values are given by

$$K_2(\infty) = \frac{\pi^2}{8}; \qquad K_3(\infty) = \frac{7\zeta(3)}{4}; \qquad K_4(\infty) = \frac{\pi^4}{16}; \qquad \ldots$$

From these, one can evaluate the limit of the skewness $\gamma_{\mathcal{T}}(\alpha)$ and kurtosis $\kappa_{\mathcal{T}}(\alpha)$ of the distribution of the absorption time as $\alpha \to \infty$ [1]:

$$\gamma_{\mathcal{T}}(\infty) = \frac{28\sqrt{2}\zeta(3)}{\pi^3} = 1.535\ldots; \qquad \kappa_{\mathcal{T}}(\infty) = 7.$$

The simplicity of the latter result compared to the former is amusing.

**3. Moments relating to the particle's position**

The form of (6) implies that the particle is overwhelmingly likely to be absorbed in a fixed time 'window' around $\mathcal{T} \approx \log \alpha$ as $\alpha \to \infty$. It is instructive to consider the corresponding position density and its associated moments when $t = \log \alpha + O(1)$. An expression for $p(x,t|\alpha)$ on $[0,\infty)$, i.e. the solution of (2), can be constructed from the well-known solution on $(-\infty,\infty)$ using the method of images:

$$p(x,t|\alpha) = \frac{1}{\sqrt{2\pi(1-e^{-2t})}} \exp\left\{-\frac{(x-\alpha e^{-t})^2}{2(1-e^{-2t})}\right\}$$

$$- \frac{1}{\sqrt{2\pi(1-e^{-2t})}} \exp\left\{-\frac{(x+\alpha e^{-t})^2}{2(1-e^{-2t})}\right\}. \tag{15}$$

As discussed earlier in relation to (3), this density is not normalised. One can construct a fully normalised density by adding a discrete contribution located at $x = 0$ to represent the absorbed sample paths:

$$(1 - E_t[1|\alpha]) \times \delta(x)$$

but for the calculation of the position moments $E_t[x^n|\alpha] = \int_0^\infty x^n p(x,t|\alpha) dx$ for $n > 0$ the distinction is immaterial. By way of example, for the first and third moments:

$$E_t[x|\alpha] = \alpha e^{-t}; \qquad E_t[x^3|\alpha] = 3\alpha e^{-t} + (\alpha^3 - 3\alpha) e^{-3t}$$

whilst for the second moment:

$$E_t[x^2|\alpha] = (1 - e^{-2t} + \alpha^2 e^{-2t}) \operatorname{erf}\left(\frac{\alpha e^{-t}}{\sqrt{2(1-e^{-2t})}}\right)$$

$$+ \sqrt{\frac{2(1-e^{-2t})}{\pi}} \alpha e^{-t} \exp\left(-\frac{\alpha^2 e^{-2t}}{2(1-e^{-2t})}\right).$$

If one sets $t = \log \alpha + \eta$, where $\eta$ is fixed, then one can show from (15) that the position density has a well-defined limiting form:

$$f(x,\eta) \equiv \lim_{\alpha \to \infty} p(x, t = \log\alpha + \eta|\alpha) = \sqrt{\frac{2}{\pi}} e^{-\frac{1}{2}e^{-2\eta}} e^{-\frac{1}{2}x^2} \sinh(xe^{-\eta}). \qquad (16)$$

The speed of convergence is illustrated in figure 2. An important check is provided by the simple observation that all sample paths which survive up to time $t = \log\alpha + \eta$ will have an absorption time characterised by $\mathcal{T} = \log\alpha + \xi$, where $\xi > \eta$. Thus

$$\Pr(\xi > \eta) = \int_0^\infty f(x,\eta)\,dx = \mathrm{erf}\left(\frac{e^{-\eta}}{\sqrt{2}}\right)$$

from which one can rederive (6); i.e. $P(\xi) = -\partial/\partial\eta \Pr(\xi > \eta)\big|_{\eta=\xi}$. Another check comes from recognising that all sample paths which survive up to time $t = \log\alpha + \eta$ may be considered to start afresh before reaching zero after a further time $\xi - \eta$. This point of view leads to the integral equation

$$\int_0^\infty f(x,\eta)\,P(\mathcal{T} = \xi - \eta | x)\,dx = P(\xi); \qquad \xi > \eta.$$

Using (4) and (16), one can prove that this holds after appropriate algebraic manipulations.

From (16), one can calculate the limiting position moments when $t = \log\alpha + \eta$, which are defined by

$$\mathrm{E}^{(\eta)}[x^n] \equiv \lim_{\alpha\to\infty} \mathrm{E}_{t=\log\alpha+\eta}[x^n|\alpha] = \int_0^\infty x^n f(x,\eta)\,dx.$$

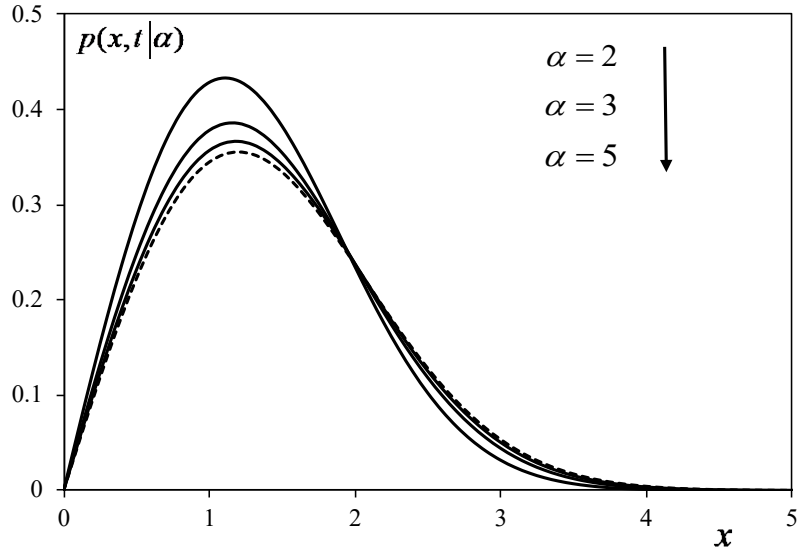

Figure 2. The position density evaluated at $t = \log \alpha$ for the indicated values of $\alpha$ (solid lines). The limit as $\alpha \to \infty$ is $f(x, \eta = 0)$ (dashed line).

The limiting moments at $t = \mathcal{T}_{\max} = \log \alpha + o(1)$ are obtained by setting $\eta = 0$. After repeated integration by parts one can show that

$$E^{(0)}[x^{2n+1}] = a_{2n+1}$$

$$E^{(0)}[x^{2n}] = a_{2n}\operatorname{erf}\left(\frac{1}{\sqrt{2}}\right) + b_{2n}\sqrt{\frac{2}{\pi e}}$$

(17)

where $a_n, b_n$ are integers which satisfy the following recursions:

$$a_n = a_{n-1} + (n-1)a_{n-2}; \qquad a_0 = 1, \quad a_1 = 1$$

$$b_n = b_{n-1} + (n-1)b_{n-2}; \qquad b_0 = 0, \quad b_1 = 1.$$

Thus $a_n = 1,1,2,4,10,26,76\ldots$ and $b_n = 0,1,1,3,6,18,48,\ldots$, although finding explicit closed-form solutions is something of a challenge. A mechanical approach utilises exhaustive integration by parts of the defining relation

$$\sqrt{\frac{2}{\pi e}}\int_0^\infty x^n e^{-\frac{1}{2}x^2} e^x \,dx \equiv a_n\left(1+\operatorname{erf}\left(\frac{1}{\sqrt{2}}\right)\right) + b_n\sqrt{\frac{2}{\pi e}}$$

and this eventually yields:

$$a_n = \sum_{k=0}^{\lfloor n/2 \rfloor} \binom{n}{2k}(2k-1)!! \tag{18}$$

$$b_n = \sum_{k=0}^{\lfloor (n-1)/2 \rfloor}\binom{n}{2k+1}\left\{\sum_{r=0}^{k}\frac{(2k)!!}{(2r)!!}\right\} - \sum_{k=0}^{\lfloor (n-2)/2 \rfloor}\binom{n}{2k+2}\left\{\sum_{r=0}^{k}\frac{(2k+1)!!}{(2r+1)!!}\right\}. \tag{19}$$

Interestingly, it turns out that the former sequence $a_n$ is widely encountered in combinatoric problems and features prominently in [12], where (18) is given. It enumerates, for example, the involution (or telephone) numbers, as well as the number of standard Young tableaux with $n$ cells. The latter sequence $b_n$ is also listed in [12] but has a much lower prominence and (19) appears to be new. For completeness, one has from (17):

$$E^{(0)}[x] = 1; \qquad E^{(0)}[x^2] = 2\text{erf}\left(\frac{1}{\sqrt{2}}\right) + \sqrt{\frac{2}{\pi e}} = 1.849...$$

$$E^{(0)}[x^3] = 4; \qquad E^{(0)}[x^4] = 10\text{erf}\left(\frac{1}{\sqrt{2}}\right) + 6\sqrt{\frac{2}{\pi e}} = 9.730...$$

(20)

which implies that, in the limit $\alpha \to \infty$, the distribution of the particle's position evaluated at the most probable absorption time has skewness $\gamma_x(\infty) = 0.577...$ and kurtosis $\kappa_x(\infty) = 2.532...$.

**4. Concluding remarks**

The essence of the results given above is as follows. When $\alpha$ is large, the evolution of $x(t)$ is initially dominated by the relaxation term in (1) and a typical sample path will remain close to the mean path $\bar{x}(t) = E_t[x|\alpha] = \alpha e^{-t}$. The noise term becomes comparable to the relaxation term for times $t = O(\log \alpha)$, at which point $x(t) = O(1)$ and, as evidenced by the position density, the particle has effectively lost memory of its initial position. One therefore has an invariant structure for the translated time when absorption occurs, which in turn is characterised by universal statistical quantities. These have been calculated explicitly.

Evidence for such invariant behaviour may be found at smaller values of $\alpha$ as well. Thus, consider the 'crossover' timescale $t_x(\alpha) = \frac{1}{2}\log(\alpha^2 + 1)$, which has the characteristic that the coefficient of variation $C_V \ll 1$ when $t \ll t_x$ and $C_V \gg 1$ when $t \gg t_x$ (for clarity, $C_V$ is the ratio of the standard deviation to the mean of the particle's position at a given time). As an immediate consequence of (3), it follows that for *any* value of the particle's initial position:

$$\Pr(\mathcal{T} > t_x) = \mathrm{erf}\left(\frac{1}{\sqrt{2}}\right) = 0.682...$$

$$\Pr(\mathcal{T} < t_x) = 1 - \mathrm{erf}\left(\frac{1}{\sqrt{2}}\right) = 0.327....$$

Moreover, from (15, 17) one has

$$p(x, t = t_x | \alpha) = \sqrt{\frac{2}{\pi e}} \frac{1}{\lambda} \exp\left(-\frac{x^2}{2\lambda^2}\right) \sinh\left(\frac{x}{\lambda}\right); \qquad \lambda \equiv \frac{\alpha}{\sqrt{\alpha^2 + 1}}$$

such that in turn

$$E_{t=t_x}[x^{2n+1} | \alpha] = \lambda^{2n+1} a_{2n+1}$$

$$E_{t=t_x}[x^{2n} | \alpha] = \lambda^{2n} \left( a_{2n} \mathrm{erf}\left(\frac{1}{\sqrt{2}}\right) + b_{2n} \sqrt{\frac{2}{\pi e}} \right)$$

(21)

with $a_n, b_n$ the same as given previously. For $t = t_x$, the coefficient of variation $C_V = 0.921...$, independent of $\alpha$. Additionally, for $t = t_x$ the distribution of the particle's position has skewness $\gamma_x = 0.577...$ and kurtosis $\kappa_x = 2.532...$, again independent of $\alpha$. In the limit $\alpha \to \infty$ one has $t_x(\alpha) = \log \alpha + o(1)$ and $\lambda = 1 + o(1)$, whereupon the earlier results are recovered.

**References**

[1]   Kearney M J and Martin R J 2021 J. Phys. A: Math. Theor. **54** 055002

[2]   Sato S 1978 Math. Biosci. **39** 53

[3]   Ricciardi L M and Sato S 1988 J. Appl. Probab. **25** 43

[4]   Alili L, Patie P and Pedersen J L 2005 Stoch. Models **4** 967

[5]   Ricciardi L M and Sacerdote L 1979 Biol. Cyber. **35** 1

[6]   Cerbone G, Ricciardi L M and Sacerdote L 1981 Int. J. Cyber. Sys. **12** 395

[7]   Nobile A G, Ricciardi L M and Sacerdote L 1985 J. Appl. Prob. **22** 360

[8]   Veestraeten D 2015 J. Appl. Probab. **52** 595

[9]   Martin R J, Kearney M J and Craster R V 2019 J. Phys. A: Math. Theor. **52** 134001

[10]  Lipton A and Kaushansky V 2020 Quant. Finance **20** 723

[11]  Giorgini L T, Moon W and Wettlaufer J S 2020 J. Stat. Phys. **181** 2404

[12]  Sequences A000085 and A000932 in *The On-Line Encyclopedia of Integer Sequences* (Editor: Sloane N J A) oeis.org